\documentclass[preprint,12pt,authoryear]{elsarticle}

\graphicspath{{./}}

\usepackage{amssymb}
\usepackage{amsmath}
\usepackage{booktabs}
\usepackage{multirow}
\usepackage{caption}
\usepackage{float}
\usepackage{xcolor}
\usepackage{xspace}
\usepackage{xurl}

\newcommand{\tool}{AutoRPG\xspace}

\newcommand{\rqhead}[1]{\elsparagraph*{#1}}

\journal{International Journal of Human-Computer Studies}

\begin{document}

\begin{frontmatter}

\title{\tool: Harnessing Systematic AI for Non-Intrusive Low-Latency Mobile RPG Combat}

\author[1]{Haibo Lin}
\author[2]{Zhong Li\corref{cor1}}
\ead{lizhong@nju.edu.cn}
\author[3]{Chunbo Qin}
\author[3]{Shu Zhang}
\author[1]{Tian Zhang}
\author[1]{Xuandong Li}

\affiliation[1]{organization={School of Computer Science, Nanjing University},
            city={Nanjing},
            state={Jiangsu},
            country={China}}
\affiliation[2]{organization={Software Institute, Nanjing University},
            city={Nanjing},
            state={Jiangsu},
            country={China}}
\affiliation[3]{organization={Huawei Technologies Co., Ltd. Nanjing},
            city={Nanjing},
            state={Jiangsu},
            country={China}}
\cortext[cor1]{Corresponding author.}

\begin{abstract}
Automated combat in mobile RPGs plays a critical role in game quality assurance, requiring real-time behavior control that closely resembles human gameplay. However, existing approaches fail to satisfy these requirements, either because they lack fine-grained behavioral control or because they incur high inference latency. To address these challenges, we incorporate user gameplay experience to design \tool, a systematic AI framework for automating mobile RPG combat. At a high level, \tool decomposes the combat process into two stages: a high-level strategic planning phase that formulates the overall combat strategy, followed by a systematic execution phase powered by a suite of carefully designed lightweight modules encompassing target tracking, character movement, dodge maneuvering, and abnormal-state handling. Experiments on three popular mobile RPGs, i.e., Genshin Impact, Wuthering Waves, and Arknights: Endfield, demonstrate that \tool consistently outperforms four competitive baselines across both combat quality and generalization capability, while behaving the most similarly to human players among all evaluated agents.
\end{abstract}

\begin{keyword}
Mobile RPG games \sep Game quality assurance \sep Automated combat \sep Vision-guided control \sep Human-like behavior \sep Low-latency inference

\end{keyword}

\end{frontmatter}

\section{Introduction}
\label{sec:intro}

The widespread popularity of mobile Role-Playing Games (RPGs) has necessitated the advancement of automated mobile RPG combat harnesses for game quality assurance~\citep{intro1, intro2, qa1, qa2, qa3, qa4, qa5}.
Automated mobile RPG combat aims to generate realistic combat executions from the visual observations and touch controls available to human players, without requiring any access to game internals or device-level modification~\citep{dqn}.
This setting is important for mobile game quality assurance because measurements are taken on-device, the harness must drive the device the same way a player would, otherwise the measurements do not reflect real gameplay~\citep{performance}.

\begin{figure}[htbp]
    \centering
    \includegraphics[width=0.8\textwidth]{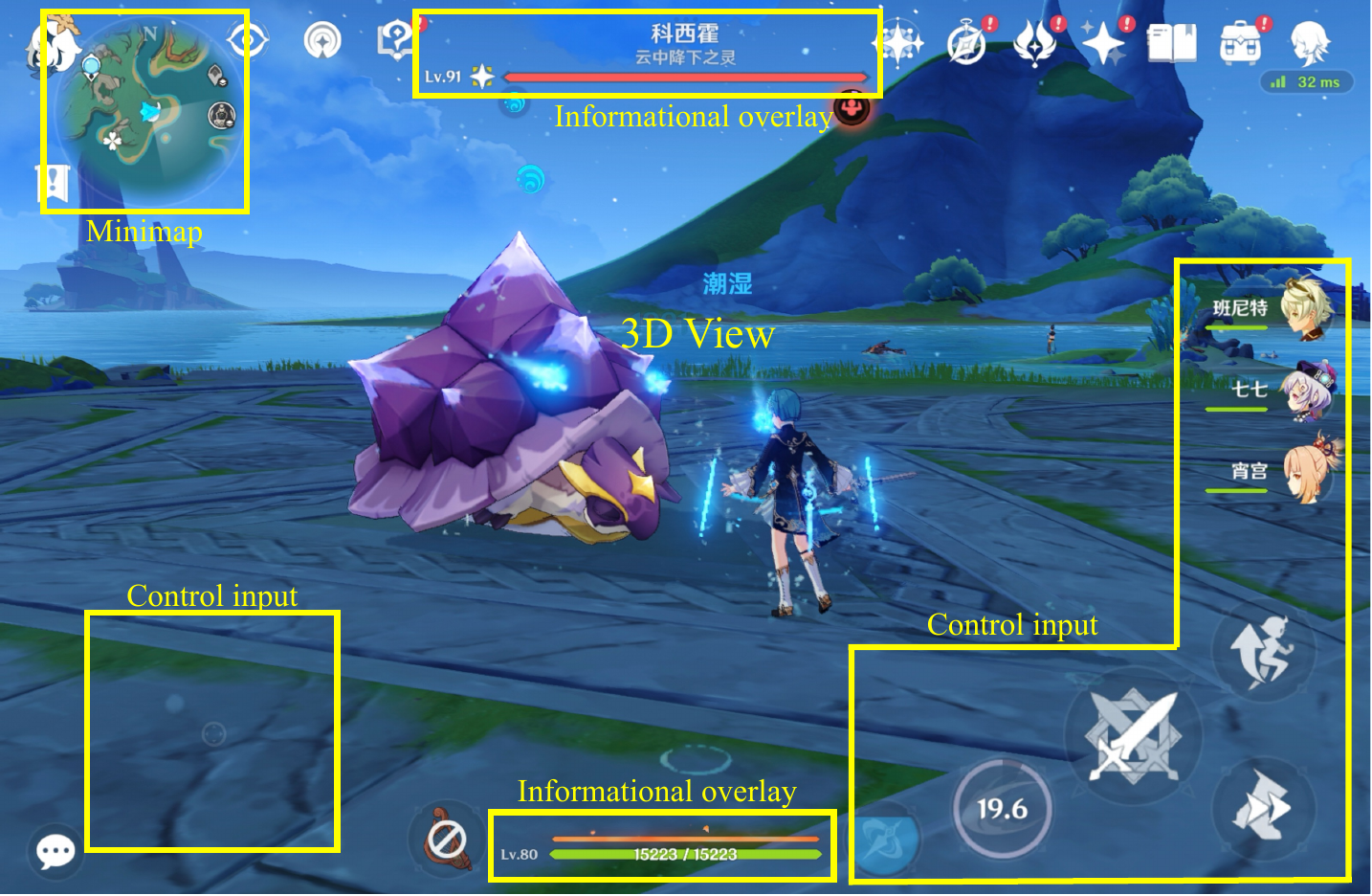}
    \caption{User interface of role-play game Genshin Impact.}
    \label{fig:interface}
\end{figure}

Unlike static UI execution in traditional software, RPG combat requires continuous perception and control in a visually complex 3D scene, engaging in combat against monsters within combat zone: the game interface integrates a centralized 3D environment view with a minimap and real-time orientation data in the top-left corner, informational overlays along the upper and lower margins, and touch controls distributed across the bottom-left and bottom-right corners of the display as shown in Figure~\ref{fig:interface}.

Within this multi-layered interface, the harness must observe combat environment  continuously from the vision followed by the execution of actions such as basic attacks, skill casting, moving, and dodging.
Combat effectiveness is jointly determined by character selection and skill-execution order, while the precision of movement and dodging governs ultimate success of the combat.
The global objective is to maximize damage output and eliminate the target while maintaining character survival through optimized action sequences.
Therefore, a practical RPG combat harness must simultaneously support strategic planning, real-time action execution, precise control, and low interference with the mobile device.

Existing game agents do not fully satisfy these requirements because their assumptions do not match real-time 3D RPG quality assurance.
Some methods rely on internal game APIs, which makes them difficult to deploy on commercial games or transfer to similar games without such interfaces~\citep{wuji, oafive, juewu5, juewu1, juewusl, juewumc, titan}.
Some methods generate only global strategies, behavior trees, or high-level plans, but do not produce reliable low-level actions that can be directly executed through the game interface~\citep{tig}.
Other methods are designed for 2D games~\citep{dqn}, FPS games~\citep{portal, mainvla}, exploration tasks with simple attack behaviors~\citep{vistawise, optimus2, jarvis-vla, openha}, and thus do not cover the behaviors required by complex 3D RPG combat.
Even among RPG-oriented methods, several systems either suffer from long inference time during combat, require large-scale training datasets, or show limited ability on difficult combat scenarios~\citep{cradle, varp, uitars, gametars, combatvla, lumine, sima2}. Table~\ref{tab:method-capability} summarizes the central gap for an automated RPG combat harness.

\begin{table}[H]
\centering
\caption{Capability comparison of representative SOTA automated game methods for real-time 3D RPG combat.
Exec.\ represents executable actions, No-API stands for API-free, 3D Cbt refers to 3D RPG combats, RT-Loop indicates a real-time loop, and Low-MC denotes low migration cost.
``\checkmark'' denotes support and ``\(\times\)'' denotes that the capability is not a primary target of the method.}
\label{tab:method-capability}
\footnotesize
\setlength{\tabcolsep}{1.5 pt}
\begin{tabular}{lccccc}
    \toprule
    \textbf{Method} & \textbf{Exec.} & \textbf{API-free} & \textbf{3D Cbt} & \textbf{RT-Loop} & \textbf{Low-MC} \\
    \midrule
    BetterGI~\citep{bgi}      & \(\checkmark\) & \(\checkmark\) & \(\checkmark\) & \(\checkmark\) & \(\checkmark\) \\
    Cradle~\citep{cradle}          & \(\checkmark\) & \(\checkmark\) & \(\checkmark\)    & \(\times\)    & \(\times\) \\
    VARP~\citep{varp}              & \(\checkmark\) & \(\checkmark\) & \(\checkmark\) & \(\times\) & \(\times\) \\
    VistaWise~\citep{vistawise}    & \(\checkmark\) & \(\checkmark\) & \(\times\)    & \(\times\) & \(\times\) \\
    DQN~\citep{dqn}                & \(\checkmark\) & \(\checkmark\) & \(\times\)    & \(\checkmark\) & \(\times\) \\
    Wuji~\citep{wuji}              & \(\checkmark\) & \(\times\) & \(\times\)    & \(\checkmark\) & \(\times\) \\
    OpenAI Five~\citep{oafive} & \(\checkmark\) & \(\times\)    & \(\times\)    & \(\checkmark\) & \(\times\) \\
    JueWu~\citep{juewu5,juewu1,juewusl} & \(\checkmark\) & \(\times\) & \(\times\) & \(\checkmark\) & \(\times\) \\
    JueWu-MC~\citep{juewumc}     & \(\checkmark\) & \(\times\) & \(\times\)    & \(\checkmark\) & \(\times\) \\
    UI-TARS~\citep{uitars}        & \(\checkmark\) & \(\checkmark\) & \(\checkmark\)    & \(\times\) & \(\times\) \\
    Game-TARS~\citep{gametars}     & \(\checkmark\) & \(\checkmark\) & \(\checkmark\) & \(\times\) & \(\times\) \\
    Optimus-2~\citep{optimus2}     & \(\checkmark\) & \(\checkmark\) & \(\times\)    & \(\times\) & \(\times\) \\
    CombatVLA~\citep{combatvla}    & \(\checkmark\) & \(\checkmark\) & \(\checkmark\) & \(\times\) & \(\times\) \\
    PORTAL~\citep{portal}          & \(\checkmark\) & \(\checkmark\) & \(\times\)    & \(\times\) & \(\times\) \\
    JARVIS-VLA~\citep{jarvis-vla}   & \(\checkmark\) & \(\checkmark\) & \(\times\)    & \(\times\) & \(\times\) \\
    TiG~\citep{tig}                & \(\times\) & \(\checkmark\) & \(\times\)    & \(\times\) & \(\times\) \\
    TITAN~\citep{titan}            & \(\checkmark\) & \(\times\) & \(\checkmark\)    & \(\times\) & \(\times\) \\
    OpenHA~\citep{openha}          & \(\checkmark\) & \(\checkmark\) & \(\times\)    & \(\times\) & \(\times\) \\
    Lumine~\citep{lumine}          & \(\checkmark\) & \(\checkmark\) & \(\checkmark\) & \(\times\) & \(\times\) \\
    SIMA 2~\citep{sima2}           & \(\checkmark\) & \(\checkmark\) & \(\checkmark\) & \(\times\) & \(\times\) \\
    NitroGen~\citep{nitrogen}      & \(\checkmark\) & \(\checkmark\) & \(\checkmark\) & \(\checkmark\) & \(\times\) \\
    MAIN-VLA~\citep{mainvla}      & \(\checkmark\) & \(\checkmark\) & \(\times\) & \(\times\) & \(\times\) \\
    \midrule
    \tool                         & \(\checkmark\) & \(\checkmark\) & \(\checkmark\) & \(\checkmark\) & \(\checkmark\) \\
    \bottomrule
\end{tabular}
\end{table}

We propose \tool, a non-intrusive, two-phase RPG combat harness that achieves precise control with low latency in the lightweight execution loop.
\tool operates exclusively through the visual game stream captured at the device display and standard touch events, the same interface available to a human player, and requires no game API, internal state hook, or device modification.
Before combat, \tool uses a large language model to select characters and design a skill-execution combo from the unlocked character set and upcoming monster information; placing this knowledge-intensive reasoning before the real-time loop ensures it does not introduce inference latency during combat.
During combat, \tool executes combat through five lightweight vision-guided modules for low latency that track the monster, attack with the planned combo, move toward the target and combat-zone center, dodge monster skills, and handle abnormal states, each specialized to precisely control one part of the combat gameplay.

We evaluate \tool on Genshin Impact, Wuthering Waves, and Arknights: Endfield from four perspectives: combat effectiveness and efficiency, human-likeness and mobile-device performance impact for automated mobile game quality assurance, cross-game generalization, and module ablations.
The results show that \tool improves combat success, damage output, and combat duty cycle over representative baselines, behaves the most similarly to human players among all evaluated agents while maintaining mobile performance measurements close to manual play, and generalizes to two additional RPG environments.
Ablation studies further show that tracking, attacking, moving, dodging, and handling each contribute to the harness's combat ability.

Our contributions are summarized as follows.
\begin{itemize}
    \item First, we formulate real-time 3D RPG combat as an automated mobile game testing problem that requires realistic, low-interference combat workload generation.
    \item Second, we propose \tool, a two-phase combat harness that combines pre-combat LLM planning with lightweight in-combat visual control.
    \item Third, we design a modular execution pipeline for tracking, attacking, moving, dodging, and handling complex RPG combat states.
    \item Fourth, we evaluate \tool across combat performance, device-performance impact, cross-game generalization, and module-level necessity.
\end{itemize}

\section{Related works}

\subsection{Embodied AI}

Embodied AI studies agents that perceive visual environments, reason about goals, and execute actions through continuous interaction. Vision-language-action (VLA) models are a representative direction in this area. RT-1~\citep{rt1} and RT-2~\citep{rt2} show that transformer policies trained on large robot datasets, and co-trained with vision-language data, can improve generalization in real-world control. OpenVLA~\citep{openvla} further provides an open-source VLA model trained on large-scale robot demonstrations. Recent studies improve the reasoning and control ability of embodied agents: ECoT~\citep{ecot} trains policies to reason about plans and visually grounded features before acting, Hume~\citep{hume} introduces value-guided System-2 thinking for robot control, and LeVERB~\citep{leverb} separates high-level vision-language reasoning from low-level whole-body control. These methods demonstrate the potential of embodied models for connecting perception, reasoning, and action, but their large-model inference cost and robotics-oriented action spaces are not directly suitable for latency-sensitive RPG combat.

\subsection{3D game agents}

3D game agents have been developed through both training-based and training-free paradigms. Training-based methods such as DQN~\citep{dqn}, OpenAI Five~\citep{oafive}, and the JueWu series~\citep{juewu5,juewu1,juewusl} and JueWu-MC~\citep{juewumc} demonstrate strong performance in Atari, Dota 2, Honor of Kings, and Minecraft through deep reinforcement learning, supervised learning, hierarchical control, or self-play. Recent agent models further scale training with screenshots, language reasoning, generated policies, or large gameplay datasets: UI-TARS~\citep{uitars} trains a native GUI agent for screenshot-based interaction, PORTAL~\citep{portal} uses LLM-generated behavior-tree policies, TiG~\citep{tig} refines language-model policies through reinforcement learning, and Game-TARS~\citep{gametars} and NitroGen~\citep{nitrogen} train generalist game agents from large-scale multimodal or gameplay data. Minecraft agents such as JARVIS-VLA~\citep{jarvis-vla}, Optimus-2~\citep{optimus2}, and OpenHA~\citep{openha} combine planning, visual perception, and low-level control for open-world tasks. Combat-oriented and virtual-world agents such as CombatVLA~\citep{combatvla}, SIMA 2~\citep{sima2}, and Lumine~\citep{lumine} further show that trained foundation agents can support game behavior, long-horizon interaction, and cross-game generalization, while automated game testing systems such as Wuji~\citep{wuji} and TITAN~\citep{titan} use learned or LLM-driven agents to improve bug discovery and task coverage. These methods can learn powerful specialized policies, but they often require large-scale data and specialized environments. For automated mobile RPG combat in game quality assurance, this migration cost is undesirable because combat scenarios frequently vary across games, monsters, and characters.

Training-free agents reduce migration cost by using foundation models or domain knowledge without retraining a full game policy. Cradle~\citep{cradle} operates software and games through screenshots and keyboard or mouse actions without relying on internal APIs. VARP~\citep{varp} studies whether VLMs can play action role-playing games from visual input and uses action planning with visual trajectory modules, while VistaWise~\citep{vistawise} integrates cross-modal domain knowledge and visual analysis to reduce data requirements in Minecraft. However, directly using large foundation models inside the combat loop introduces high inference latency and coarse action precision, which are problematic for fast RPG combat. Compared with both training-based and training-free agents, \tool focuses on a practical requirement: low-latency, RPG combat harness with precise control for game quality assurance. Its two-phase design bridges the gap by using LLM planning for pre-combat character selection and combo design, while using lightweight tracking, attacking, moving, dodging, and handling modules for precise real-time combat execution.

\section{Method}
\label{sec:method}

\begin{figure}[htbp]
    \centering
    \includegraphics[width=\textwidth]{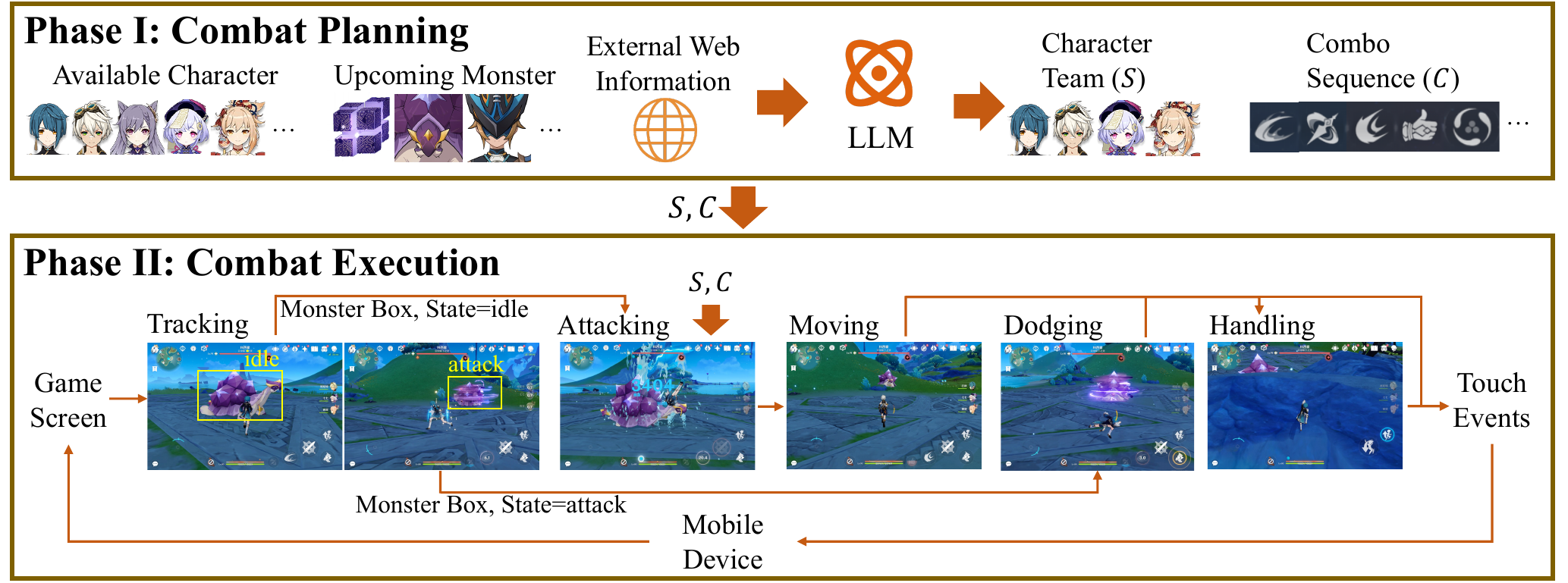}
    \caption{\tool: a non-intrusive, two-phase combat harness for precise control in real-time RPG combat with low latency.}
    \label{fig:overview}
\end{figure}

The design of the combat harness was informed by semi-structured interviews with experienced RPG players.
Interviewees described a separation between planning and execution of combat.
Players usually invest significant effort in choosing the optimal character team for the target monster and selecting a skill-execution order that maximizes damage output, reasoning about monster attributes, available character pool, and the character abilities.
Critically, this planning does not need to happen during combat: once a strategy is decided, players can execute it throughout the fight with other necessary actions.
The separation between pre-combat reasoning and in-combat action motivates \tool's two-phase architecture.
Figure~\ref{fig:overview} shows the overview of \tool, a non-intrusive, two-phase combat harness for precise control in real-time RPG combat with low latency.
Before combat, \tool performs combat planning to select characters and design a combo for the upcoming monster.
During combat, \tool performs combat execution through five lightweight vision-guided modules: tracking, attacking, moving, dodging, and handling.
This separation allows \tool to use strong language-model reasoning where latency is tolerable, while keeping the measured combat loop fast enough for mobile game quality assurance.

\subsection{Combat planning}

Combat planning converts scenario information into an executable combat strategy before the real-time loop starts.
RPG combat depends heavily on character selection and skill order, but these decisions require semantic knowledge about character abilities, monster properties, and damage synergy.
However, the combinatorial complexity in characters and the myriad of possible skill sequences necessitates a planning component capable of sophisticated reasoning.
We propose LLM as the reasoning engine, because the planner must combine character abilities, monster attributes, and external strategy guides; the kind of multi-source semantic synthesis LLMs do better than rule-based planners.
Since large language model inference is too expensive to run at every combat step, \tool invokes the planner only before combat begins, inspired by the interviewees of the pre-combat planning.
Given the unlocked character information \(I_C\) and the upcoming monster information \(I_M\), \tool constructs a prompt \(P(I_C, I_M)\) and allows the planning model \(\mathcal{M}_P\) to incorporate relevant external knowledge \(I_{ext}\) through web search.
The planner generates
\[
    \mathcal{M}_P(P(I_C, I_M), I_{ext}) \to \{S, C\}
\]
where \(S\) is the selected character team and \(C\) is the planned combo sequence.
\tool switches to the generated characters \(S\) before combat, and passes \(C\) to the attacking module as a reusable high-level action plan.
This design makes the strategy adaptable to different monsters and character pools without placing LLM inference inside the latency-critical combat loop.

\subsection{Combat execution}

Combat execution maintains a realistic and continuous combat workload through vision-guided control with visual observation and touch controls following gameplay by human players.
A human player's combat in mobile RPG follows five parts until the monster is defeated:
(1) Track: The player reads the 3D scene on every frame, tracking the monster's on-screen position and watching for the start of an attack animation.
(2) Attack: When the monster is idle, the player executes the pre-planned skills rotation to deal damage on the monster.
(3) Move: When the character has been away from the monster or displaced from the combat zone, the player periodically closes the gap to the monster or backs toward the combat zone so that subsequent attacks can hit and the encounter does not reset.
(4) Dodge: When the monster begins an attack, the player performs an evasion, a quick lateral step for simple attacks, or a long evasion trajectory for complex multi-directional attacks.
(5) Handle: When an environmental effect locks the character into an abnormal state like climbing, freezing, or swimming, the player applies a state-specific action to break out and resume normal combat.
The global objective is to maximize damage output and eliminate the monster while maintaining character survival.
\tool's combat execution phase mirrors these five parts through five lightweight modules, each designed to automate exactly one part of the human gameplay.
\tool observes the current game frame to determine whether the character should track, attack, move, dodge, or handle an abnormal state, and execute control by standard touch events.
The execution phase is intentionally modular to align with the core mechanics and dynamic requirements of RPG combat.
This specialized structure allows the agent to navigate several distinct operational demands with precise control in the RPG combat.
To ensure real-time efficiency, \tool combines computer vision algorithms, lightweight neural networks, and a small VLA model only for complex dodging.

\paragraph{Tracking}
The tracking module is the shared perception backbone whose outputs every downstream module depends on.
Real-time RPG combat control depends on continuously observing both where the monster appears in the frame and whether it is currently attacking.
To supply this information, tracking module runs a YOLO-based detector~\citep{yolo} on each game frame to produce a monster bounding box and a state label.
The bounding box center is consumed by the moving module to estimate proximity, and the state label is consumed by the attacking and dodging modules to determine their respective actions.
Due to the monster drifting to the frame periphery as the character or monster move, the tracking module recenters the target by rotating the in-game camera: it converts the bounding box's horizontal offset from the frame center into horizontal drag gestures on an empty screen area, which the game interprets as camera yaw rotation, keeping the bounding box near the center of the field of view and within a reliable detection region on every subsequent frame.

\paragraph{Attacking}
The attacking module turns the pre-combat designed combo into reproducible damage output.
Using the fixed pre-planned combo follows the gameplay of human players described in interviews.
When the tracking module indicates that the monster is not attacking, the attacking module sequentially executes the planned combo \(C\).

\paragraph{Moving}
The moving module addresses two spatial displacement problems: the character being outside the combat zone, and drifting beyond effective attack range for hit the monster, which caused by attacks from or on characters, as well as dodging.
To recover from combat-zone displacement, the module reads the minimap and applies image matching to estimate the character's position to the zone; a CNN model simultaneously identifies the character's orientation from the minimap.
With this position-and-orientation estimate, the module computes a return direction and controls the character move toward the combat-zone center after each attack sequence, preventing unintended combat resets caused by out-of-zone displacement.
To recover from attack-range displacement, the module uses the monster bounding box area provided by the tracking module as a proxy for character-to-monster distance: a larger box means the character is close; a smaller box means the character has drifted back.
When the bounding box area falls below a close-range threshold, the module maps the scale difference to a forward movement duration and issues an approach command that closes the gap before the next attack, reducing missed attacks caused by excessive distance.
By separating zone-level control from range-level control, the moving module maintains the spatial preconditions for every attack.

\paragraph{Dodging}
The dodging module protects the character from monster skills so that the combat remains active.
When the tracking module detects that the monster is attacking, \tool chooses the dodging policy according to attack complexity.
For simple attacks, the module performs lateral evasion with rule-based control.
For complex attacks, \tool uses a small-scale VLA model to generate continuous dodging directions from recent visual observations.
This hybrid design reserves learned visual reasoning for cases where handcrafted rules are insufficient, while keeping common evasive actions lightweight.

We train the dodging VLA model in two stages.
The first stage uses supervised fine-tuning (SFT) to teach basic visual grounding and direction generation from human annotations.
For each training instance, the model receives the most recent \(k\) frames \(F=\{f_1, f_2, \cdots, f_k\}\) and predicts an evasion trajectory \(D=\{d_1, d_2, \cdots, d_n\}\).
The second stage uses Proximal Policy Optimization (PPO) in actual combat scenarios to optimize survival-oriented behavior.
During PPO training, positive rewards encourage valid direction generation, damage avoidance, and positional stability, while penalties discourage invalid directions, damage taken, and excessive displacement.
Detailed reward function is provided in \ref{app:reward}.
This two-stage training makes the model first imitate human dodging behavior and then adapt that behavior to the survival objective of real RPG combat.

\paragraph{Handling}
The handling module recovers from exceptional states that would invalidate a combat.
When the character enters states such as climbing, freezing, or swimming, normal attack, movement, and dodging commands no longer produce useful combat behavior.
\tool therefore applies state-specific recovery logic: climbing is terminated by jumping over obstacles, freezing is resolved by rapidly clicking the ice-breaking button, and swimming is handled by swimming out of the water zone.
Abnormal states are detected via visual template matching on specific UI icons, such as the ice-breaking button for freezing and swimming and climbing are identified by specific control input pattern on the screen, without accessing the game's internal state.
This module keeps the harness from silently degrading into idle or invalid behavior during automated combat.

\section{Evaluation}

We design the evaluation to validate four claims of \tool: effectiveness and efficiency in 3D RPG combat, human-like combat behavior with low device interference for automated mobile game quality assurance, cross-game generalization, and the necessity of each execution module.
To assess the performance of \tool, we conduct experiments to address following research questions:
\begin{itemize}
    \item \textbf{RQ1:} How effective and efficient is \tool in combat?
    \item \textbf{RQ2:} How human-like is \tool's combat behavior, and how does it affect the mobile device's performance compared to manual playing?
    \item \textbf{RQ3:} How is the generalization performance of \tool?
    \item \textbf{RQ4:} How do the modules of \tool contribute to the effectiveness in combat?
\end{itemize}

\subsection{Experiment settings}
\label{sec:setup}

\paragraph{Task definition}
We employ three games, Genshin Impact, Wuthering Waves, and Arknights: Endfield, as experiment platforms given their global popularity in mobile games.
We define 10 combat tasks in Genshin Impact, 4 combat tasks in Wuthering Waves, and 4 combat tasks in Arknights: Endfield based on their difficulty levels, categorizing them into four distinct levels: easy, medium, hard, and very hard. Detailed tasks are provided in \ref{app:task}.

\paragraph{Dataset}

We collect three small-scale datasets to train the lightweight perception modules in \tool.
For the tracking module, annotated game frames with 1000 to 2750 per monster are used to fine-tune a YOLO detection model.
For the moving module, 515 minimap samples annotated for character orientation and camera direction are augmented to 2575 images for CNN training.
For the dodging module, 63 gameplay videos of continuous dodge actions are segmented into 1057 four-frame data pairs with direction labels for VLA SFT stage.
Full annotation procedures and train/valid/test splits are provided in \ref{app:dataset}.

\paragraph{Baselines}
We compare \tool against four representative baselines, as listed in what follows: (a) BetterGI~\citep{bgi}: a script-based combat automation; (b) AI-Wukong~\citep{bmwai}: a vision-based reinforcement learning approach; (c) UI-TARS~\citep{uitars}: a fine-tuned GUI agent; and (d) NitroGen~\citep{nitrogen}: a vision-action foundation model.
All baselines are adapted for mobile devices, specifically targeting Genshin Impact, Wuthering Waves, and Arknights: Endfield.
Because these methods are not designed for combat planning, all baselines use the same character team produced by \tool to isolate combat-execution ability from character-selection quality.
Methods designed for non-combat settings or unavailable as not open source are excluded.
The details of baselines and full justifications of excluded methods appear in \ref{app:baselines}.
Notably, UI-TARS requires a synchronized inference protocol that pauses the game during model inference, which \tool does not require.
This accommodation favors the baseline and thus represents a conservative comparison for \tool.

\paragraph{Implementation}

\tool uses DeepSeek-R1~\citep{deepseek} for pre-combat planning.
A scrcpy-based framework~\citep{scrcpy} is employed for real-time screen capture and multi-touch interaction on the mobile device.
Tracking module employs a fine-tuned YOLOv11-X model~\citep{yolo}.
Orientation recognition in moving module uses a CNN adapted from the pytorch\_image\_regression repository~\citep{pytorch_image_regression};
VLA model in dodging module relies on Qwen2.5-VL-3B~\citep{qwen2_5vl} fine-tuned with SFT followed by PPO.
Complete hyperparameter settings are listed in \ref{app:hyper}.

\paragraph{Experiment environment}

All training and real-time inference during combat are conducted on a workstation with an Intel Xeon Gold 6426Y, 503GB memory, and an NVIDIA RTX A6000 GPU, running Ubuntu 22.04.
On-device combat is evaluated on a Xiaomi Pad 7 Pro with 8GB of RAM.

\subsection{RQ1: Evaluating effectiveness and efficiency}

\begin{figure}[htbp]
    \centering
    \includegraphics[width=0.95\textwidth]{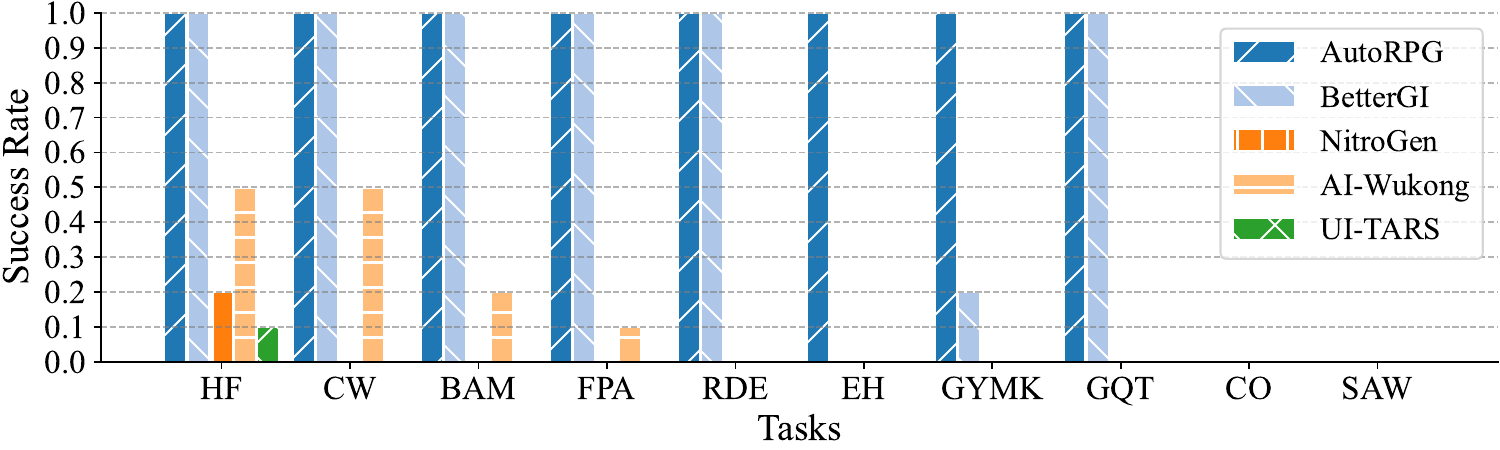}
    \caption{Results of Success Rate of \tool and baselines when executing combat tasks in Genshin Impact.}
    \label{fig:rq1_sr}
\end{figure}

\paragraph{Setup} In RQ1, we evaluate the effectiveness and efficiency of \tool. To do this, we respectively apply \tool, BetterGI, AI-Wukong, UI-TARS, and NitroGen to combat tasks in Genshin Impact listed in Table~\ref{tab:task}, with each task executed for ten independent trials. We measure their effectiveness and efficiency based on the following metrics: (1) Success Rate: the ratio of successfully defeating the monster to the total number of evaluation trials; (2) Damage Dealt: total damage output produced throughout the duration of a single combat encounter; (3) Combat Duty Cycle: the ratio of the time the harness spends executing combat to the total session duration, which includes the overhead of model inference and decision-making processes. Details of these metrics are provided in \ref{app:metrics}.

\paragraph{Results of effectiveness} Figure \ref{fig:rq1_sr} and Table~\ref{tab:rq1_dd_cd} present the results of \tool and baselines in effectiveness.
As shown in Figure~\ref{fig:rq1_sr} and Table~\ref{tab:rq1_dd_cd}, \tool outperforms the 4 baselines on all combat tasks in Genshin Impact.
Specifically, compared to the baselines, \tool achieves a Success Rate of 1.0 across easy, medium, and hard tasks, consistently matching or outperforming all established baselines.
Regarding Damage Dealt, \tool reaches the maximum value 1.0 in easy and medium tasks, performing on par with BetterGI.
\tool also reaches the maximum value 1.0 in hard tasks, surpassing BetterGI by up to 0.80, while exceeding other baselines by margins of 0.17 to 0.99, 0.65 to 1.00, and 0.93 to 1.00 across the three difficulty tasks, respectively.
Although \tool fails to complete the very hard tasks, it demonstrates superior performance in sub-metrics compared to the baselines, achieving a 0.14 to 0.26 higher Damage Dealt.
These results demonstrate that \tool can complete combat tasks more effectively.

\paragraph{Results of efficiency} Table \ref{tab:rq1_cdc} presents the results of \tool and baselines in efficiency.
According to the Table~\ref{tab:rq1_cdc}, \tool demonstrates efficiency that is either on par with or superior to the baselines.
Specifically, \tool attains a Combat Duty Cycle of 1.0 across all tasks, representing a significant improvement of 0.41 to 0.60 over UI-TARS while remaining competitive with other baselines.
These results underscore the efficiency of our approach
with low latency in combat execution.

\begin{table}[H]
    \centering
    \caption{Statistical results of Damage Dealt for combat tasks in Genshin Impact. Results are reported in the format of Mean (Standard Deviation).
For the Damage Dealt metric, values are normalized such that a full monster health bar equals 1.0.
Bold text highlights the most superior results in each row.}
    \label{tab:rq1_dd_cd}
    \footnotesize
\setlength{\tabcolsep}{4 pt}
\begin{tabular}{@{}cccccc@{}}

\toprule
\textbf{Task} & \textbf{\tool} & \textbf{BetterGI} & \textbf{NitroGen} & \textbf{AI-Wukong} & \textbf{UI-TARS} \\
\cmidrule(lr){1-6}
HF & \textbf{1.00(0.00)} & \textbf{1.00(0.00)} & 0.62(0.36) & 0.83(0.23) & 0.14(0.31) \\
CW & \textbf{1.00(0.00)} & \textbf{1.00(0.00)} & 0.03(0.03) & 0.77(0.31) & 0.01(0.03) \\
BAM & \textbf{1.00(0.00)} & \textbf{1.00(0.00)} & 0.07(0.08) & 0.35(0.36) & 0.00(0.00) \\
FPA & \textbf{1.00(0.00)} & \textbf{1.00(0.00)} & 0.04(0.05) & 0.28(0.33) & 0.04(0.05) \\
RDE & \textbf{1.00(0.00)} & \textbf{1.00(0.00)} & 0.03(0.03) & 0.11(0.08) & 0.00(0.00) \\
EH & \textbf{1.00(0.00)} & 0.20(0.28) & 0.00(0.01) & 0.01(0.02) & 0.00(0.00) \\
GYMK & \textbf{1.00(0.00)} & 0.64(0.31) & 0.01(0.01) & 0.01(0.00) & 0.00(0.00) \\
GQT & \textbf{1.00(0.00)} & \textbf{1.00(0.00)} & 0.01(0.01) & 0.07(0.04) & 0.01(0.01) \\
CO & \textbf{0.26(0.13)} & 0.07(0.04) & 0.00(0.00) & 0.01(0.01) & 0.00(0.00) \\
SAW & \textbf{0.18(0.08)} & 0.04(0.02) & 0.00(0.00) & 0.00(0.00) & 0.00(0.00) \\
\bottomrule

\end{tabular}

\end{table}
\begin{table}[H]
    \centering
    \caption{Statistical results of Combat Duty Cycle for combat tasks in Genshin Impact. Results are reported in the format of Mean (Standard Deviation). Bold text highlights the most superior results in each row.}
    \label{tab:rq1_cdc}
    \footnotesize
\setlength{\tabcolsep}{4 pt}
\begin{tabular}{@{}cccccc@{}}

\toprule
\textbf{Task} & \textbf{\tool} & \textbf{BetterGI} & \textbf{NitroGen} & \textbf{AI-Wukong} & \textbf{UI-TARS} \\
\cmidrule(lr){1-6}
\cmidrule(lr){1-6}
HF & \textbf{1.00(0.00)} & \textbf{1.00(0.00)} & \textbf{1.00(0.00)} & \textbf{1.00(0.00)} & 0.57(0.09) \\
CW & \textbf{1.00(0.00)} & \textbf{1.00(0.00)} & \textbf{1.00(0.00)} & \textbf{1.00(0.00)} & 0.57(0.08) \\
BAM & \textbf{1.00(0.00)} & \textbf{1.00(0.00)} & \textbf{1.00(0.00)} & \textbf{1.00(0.00)} & 0.59(0.04) \\
FPA & \textbf{1.00(0.00)} & \textbf{1.00(0.00)} & \textbf{1.00(0.00)} & \textbf{1.00(0.00)} & 0.55(0.06) \\
RDE & \textbf{1.00(0.00)} & \textbf{1.00(0.00)} & \textbf{1.00(0.00)} & \textbf{1.00(0.00)} & 0.44(0.10) \\
EH & \textbf{1.00(0.00)} & \textbf{1.00(0.00)} & \textbf{1.00(0.00)} & \textbf{1.00(0.00)} & 0.54(0.08) \\
GYMK & \textbf{1.00(0.00)} & \textbf{1.00(0.00)} & \textbf{1.00(0.00)} & \textbf{1.00(0.00)} & 0.40(0.06) \\
GQT & \textbf{1.00(0.00)} & \textbf{1.00(0.00)} & \textbf{1.00(0.00)} & \textbf{1.00(0.00)} & 0.40(0.03) \\
CO & \textbf{1.00(0.00)} & \textbf{1.00(0.00)} & \textbf{1.00(0.00)} & \textbf{1.00(0.00)} & 0.48(0.06) \\
SAW & \textbf{1.00(0.00)} & \textbf{1.00(0.00)} & \textbf{1.00(0.00)} & \textbf{1.00(0.00)} & 0.51(0.07) \\
\bottomrule

\end{tabular}

\end{table}

\subsection{RQ2: Evaluating automated mobile game quality assurance capability}

\begin{figure}[htbp]
\begin{minipage}[b]{0.48\textwidth}
    \centering
    \includegraphics[width=\textwidth]{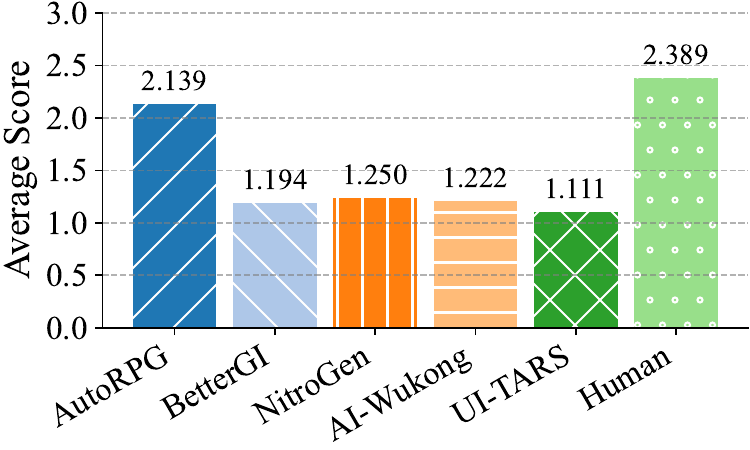}
    \captionof{figure}{Average human-likeness scores of combat replays of \tool, baselines, and human players in the user study. Each system is judged with two replays by 18 raters on a three-level scale (3: completely human operation, 2: hard to tell, 1: obviously agent operation).}
    \label{fig:rq2_humanlike}
\end{minipage}
\hfill
\begin{minipage}[b]{0.48\textwidth}
    \centering
    \includegraphics[width=\textwidth]{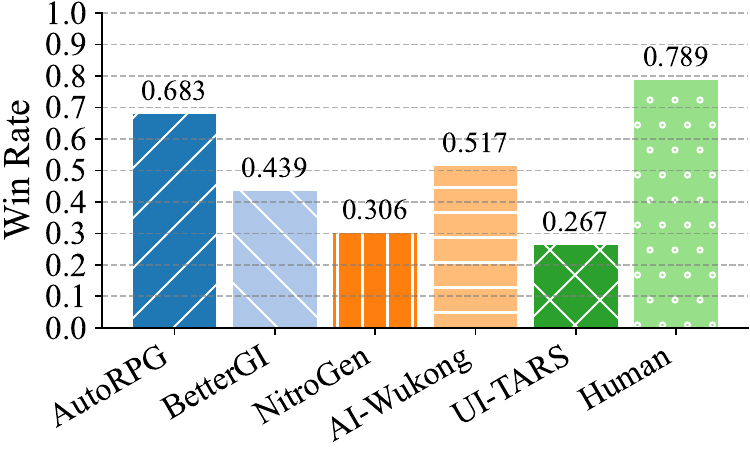}
    \captionof{figure}{Average win rates of combat replays of \tool, baselines, and human players in the pairwise preference user study. A vote for a clip scores 1, a vote for its opponent scores 0, and ``hard to tell'' scores 0.5 for both clips.}
    \label{fig:rq2_pairwise}
\end{minipage}
\end{figure}

\paragraph{Setup} In RQ2, we evaluate the capability of \tool for automated mobile game quality assurance, which requires the harness to behave like human players while imposing minimal impact on the mobile device.
To this end, we assess two aspects: the human-likeness of combat behavior and the impact on mobile device performance.

First, we conduct a user study to assess the human-likeness of combat behavior.
For each of \tool, the four baselines, and a human player, we record two combat replays on the very hard task Cocijo in Genshin Impact (12 replays in total) for human-likeness rating, and two combat replays on the hard task Electro Hypostasis for pairwise comparison.
We recruit 20 raters, each of whom (1) watches the 12 Cocijo replays and rates each replay on a three-level scale: completely human operation, hard to tell, or obviously agent operation, and (2) makes 15 pairwise preference judgments over the Electro Hypostasis replays, choosing which replay of a pair looks more like human operation, with a ``hard to tell'' option.
All raters read the study information and voluntarily consent to participate prior to the anonymous study.
Two attention-check questions are embedded in the questionnaire; two raters who failed an attention check were excluded, leaving 18 valid raters.

Second, we evaluate the impact of \tool on mobile device performance during combat.
To achieve this, we conduct ten combat trials on the very hard task Cocijo using both \tool and manual playing.
We choose Cocijo because it is a very hard task that produces sustained combat pressure and therefore stresses device performance.
We measure their performance by comparing the following key performance metrics: (1) Frames Per Second; (2) CPU Temperature; (3) GPU Temperature; (4) Battery Temperature. These metrics are reported as mean values derived from continuous data sampling during the experiments by HiSmartPerf~\citep{hisp}. Details of these metrics are provided in \ref{app:metrics}.

\paragraph{Results of human-likeness}
Figure~\ref{fig:rq2_humanlike} presents the average human-likeness scores of \tool, baselines, and human players.
As illustrated in Figure~\ref{fig:rq2_humanlike}, \tool achieves an average score of 2.14, the closest to human players (2.39) and far above the four baselines, which score only 1.11 to 1.25.
Figure~\ref{fig:rq2_pairwise} further shows that in pairwise preference, \tool attains the highest win rate among all agent systems, second only to human players.
These results demonstrate that \tool's combat behavior is substantially more human-like than existing agents and approaches that of human players.

\paragraph{Results of device performance}

Table \ref{tab:rq2} presents the results of \tool and manual playing in mobile device performance.
As illustrated in the Table~\ref{tab:rq2}, the performance impact of \tool closely aligns with that of manual playing, showing marginal differences acceptable for automated mobile game quality assurance.
Specifically, when compared to manual play, \tool exhibits a negligible reduction in FPS of only 0.03, while CPU, GPU, and battery temperatures increase by a slight margin of 5.27, 3.83, 0.7, respectively.
These results demonstrate that \tool imposes minimal computational overhead on mobile devices, ensuring its suitability for automated mobile game quality assurance.

\begin{table}[H]
    \centering
    \caption{Statistical results of mobile performance when execute combat with Cocijo in Genshin Impact. Results are reported in the format of Mean (Standard Deviation). Bold text highlights the most superior results in each row.}
    \label{tab:rq2}
    \footnotesize
\setlength{\tabcolsep}{5 pt}
\begin{tabular}{@{}ccc@{}}

\toprule
\textbf{Performance} & \textbf{\tool} & \textbf{Manual} \\
\cmidrule(lr){1-3}
Frames Per Second & 30.17(0.11) & \textbf{30.20(0.00)} \\
CPU Temperature & 59.07(1.65) & \textbf{53.80(0.73)} \\
GPU Temperature & 51.53(0.85) & \textbf{47.70(0.65)} \\
Battery Temperature & 31.17(0.61) & \textbf{30.47(0.43)} \\
\bottomrule

\end{tabular}

\end{table}
\begin{table}[H]
    \centering
    \caption{Statistical results of Damage Dealt for combat tasks in Wuthering Waves. Results are reported in the format of Mean (Standard Deviation).
For the Damage Dealt metric, values are normalized such that a full monster health bar equals 1.0.
Bold text highlights the most superior results in each row.}
    \label{tab:rq3_dd_cd}
    \footnotesize
\setlength{\tabcolsep}{4 pt}
\begin{tabular}{@{}cccccc@{}}

\toprule
\textbf{Task} & \textbf{\tool} & \textbf{BetterGI} & \textbf{NitroGen} & \textbf{AI-Wukong} & \textbf{UI-TARS} \\
\cmidrule(lr){1-6}
FNR & \textbf{1.00(0.00)} & \textbf{1.00(0.00)} & 0.27(0.14) & 0.59(0.22) & 0.04(0.06) \\
SC & \textbf{1.00(0.00)} & 0.98(0.07) & 0.15(0.11) & 0.41(0.26) & 0.01(0.02) \\
HE & \textbf{0.80(0.17)} & 0.55(0.06) & 0.13(0.10) & 0.20(0.13) & 0.00(0.01) \\
DOD & \textbf{0.51(0.22)} & 0.14(0.08) & 0.03(0.04) & 0.06(0.03) & 0.00(0.01) \\
\bottomrule

\end{tabular}

\end{table}

\subsection{RQ3: Evaluating generalization}

\paragraph{Setup} In RQ3, we evaluate the generalization of \tool. To do this, we respectively apply \tool, BetterGI, AI-Wukong, UI-TARS, and NitroGen to combat tasks in Wuthering Waves and Arknights: Endfield listed in Table~\ref{tab:task}, with each task executed for ten independent trials. We measure their generalization based on the Success Rate and Damage Dealt.

\begin{figure}[htbp]
\begin{minipage}[b]{0.48\textwidth}
    \centering
    \includegraphics[width=\textwidth]{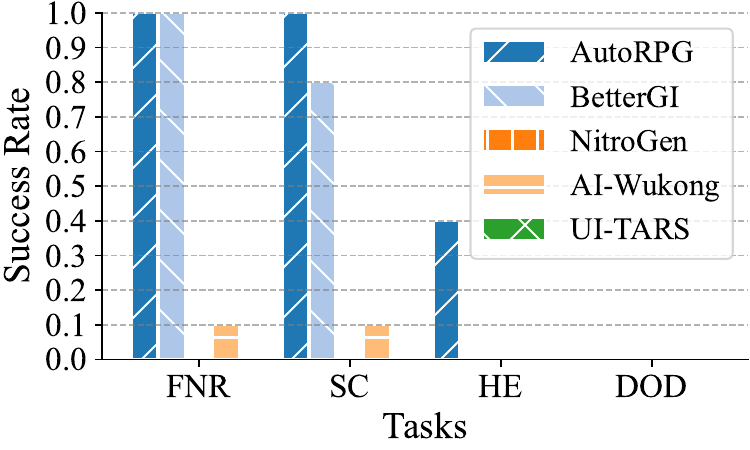}
    \captionof{figure}{Results of Success Rate of \tool and baselines when executing combat tasks in Wuthering Waves.}
    \label{fig:rq3_sr}
\end{minipage}
\hfill
\begin{minipage}[b]{0.48\textwidth}
    \centering
    \includegraphics[width=\textwidth]{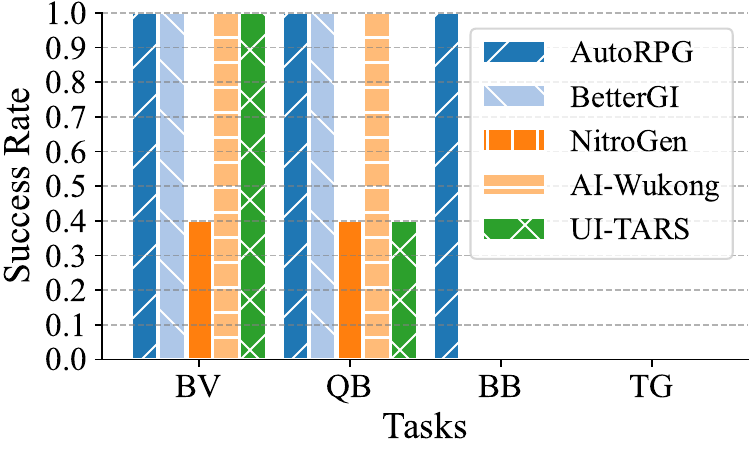}
    \captionof{figure}{Results of Success Rate of \tool and baselines when executing combat tasks in Arknights: Endfield.}
    \label{fig:rq3_endfield_sr}
\end{minipage}
\end{figure}

\begin{figure}[htbp]
    \centering
    \includegraphics[width=0.95\textwidth]{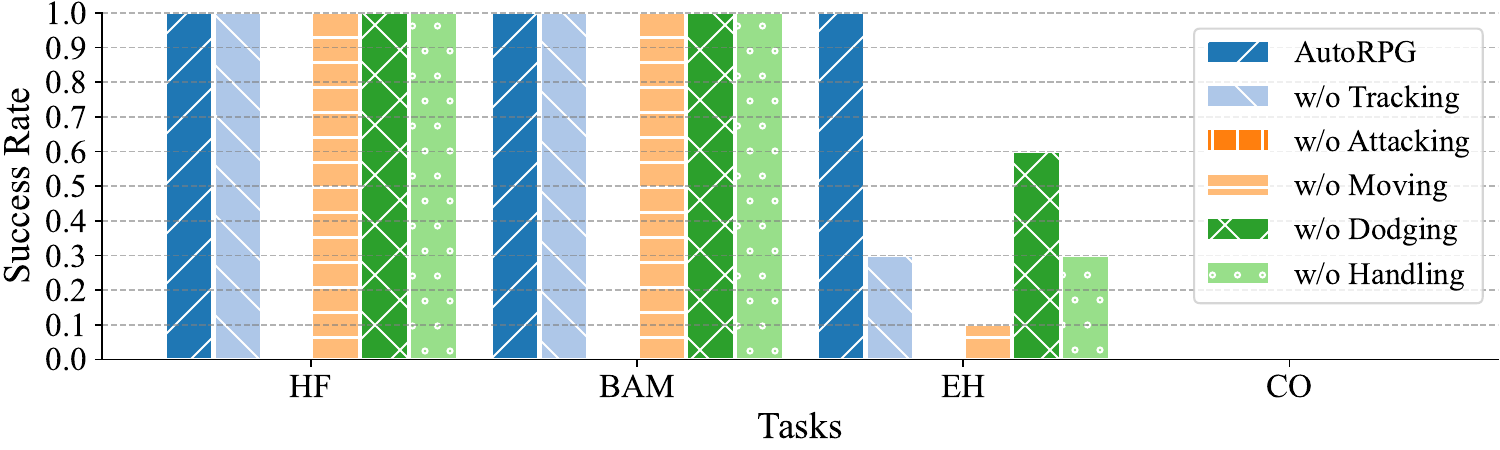}
    \caption{Results of Success Rate of \tool and ablation variants when executing combat tasks in Genshin Impact.}
    \label{fig:rq4_sr}
\end{figure}

\paragraph{Results} Figure~\ref{fig:rq3_sr} and Figure~\ref{fig:rq3_endfield_sr} present the Success Rate of \tool and baselines when executing combat tasks in Wuthering Waves and Arknights: Endfield, and Table~\ref{tab:rq3_dd_cd} and Table~\ref{tab:rq3_endfield_dd} present the Damage Dealt.
As illustrated in Figure~\ref{fig:rq3_sr} and Table~\ref{tab:rq3_dd_cd}, \tool consistently outperforms the 4 baselines across all combat tasks in Wuthering Waves.
Specifically, \tool achieves a perfect Success Rate of 1.0 in easy and medium tasks, while maintaining a 0.4 on hard task, consistently matching or outperforming all established baselines.
In terms of Damage Dealt, \tool reaches the theoretical maximum of 1.00 in easy and medium tasks while 0.80 in hard task, exceeding other baselines by margins of up to 0.96, 0.02 to 0.99 and 0.25 to 0.80 across the three difficulty tasks, respectively.
\tool sustains its superiority in the very hard task, attaining a Damage Dealt of 0.51, which exceeds the baselines by 0.37 to 0.51.

We observe a similar trend on Arknights: Endfield.
Specifically, \tool achieves a Success Rate of 1.0 in easy, medium, and hard tasks, consistently matching or outperforming all established baselines.
In terms of Damage Dealt, \tool reaches the theoretical maximum of 1.00 in easy, medium, and hard tasks, exceeding other baselines by margins of up to 0.25, up to 0.37 and 0.47 to 0.87 across the three difficulty tasks, respectively.
On the very hard task, where no method completes the task, \tool still attains the highest Damage Dealt of 0.60, exceeding the baselines by 0.11 to 0.52.
These results demonstrate that \tool can complete combat tasks more effectively while generalizing to two additional RPG environments.

\begin{table}[H]
\centering
\caption{Statistical results of Damage Dealt for combat tasks in Arknights: Endfield. Results are reported in the format of Mean (Standard Deviation).
For the Damage Dealt metric, values are normalized such that a full monster health bar equals 1.0.
Bold text highlights the most superior results in each row.}
\label{tab:rq3_endfield_dd}
\footnotesize
\setlength{\tabcolsep}{2 pt}
\begin{tabular}{@{}cccccc@{}}

\toprule
\textbf{Task} & \textbf{\tool} & \textbf{BetterGI} & \textbf{NitroGen} & \textbf{AI-Wukong} & \textbf{UI-TARS} \\
\cmidrule(lr){1-6}
BV & \textbf{1.00(0.00)} & \textbf{1.00(0.00)} & 0.75(0.34) & \textbf{1.00(0.00)} & \textbf{1.00(0.00)} \\
QB & \textbf{1.00(0.00)} & \textbf{1.00(0.00)} & 0.64(0.37) & \textbf{1.00(0.00)} & 0.63(0.33) \\
BB & \textbf{1.00(0.00)} & 0.53(0.07) & 0.29(0.08) & 0.43(0.14) & 0.13(0.04) \\
TG & \textbf{0.60(0.11)} & 0.49(0.06) & 0.20(0.08) & 0.32(0.07) & 0.08(0.02) \\
\bottomrule

\end{tabular}
\end{table}

\subsection{RQ4: Ablation study}
\label{sec:rq4}

\paragraph{Setup} In RQ4, we analyze the contribution of each module in \tool.
We construct five variants of \tool:
(1) w/o Tracking, which omits the tracking module;
(2) w/o Attacking, which omits the attacking module;
(3) w/o Moving, which omits the moving module;
(4) w/o Dodging, which omits the dodging module;
(5) w/o Handling, which omits the handling module.
We apply these variants to combat in easy task Hilichurl Fighter, medium task Blazing Axe Mitachurl, hard task Electro Hypostasis, and very hard task Cocijo, for studying the impact of these modules on effectiveness.

\paragraph{Results} Figure~\ref{fig:rq4_sr}
presents the comparison results between \tool and its variants.
From these results, we make the following observations.
First, \tool executes combat more effectively compared to the five variants, demonstrating the contribution of each module.
These modules deal each part of combat, ensuring the effectiveness of executing combat.
Second, tracking module is essential for choosing actions with ability to perceive the monster's state and relative position.
 Its removal causes a marked decline in Success Rate as the characters can no longer choose best actions without the monster state.
 Without this module, the variant fails to perceive the monster's state or relative position, leading to missed attacks and an inability to execute timely dodges, which results in lower damage and premature defeat in hard and very hard tasks.
Third, the attacking module serves as the primary driver of combat.
 Removing the attacking module causes the Success Rate to drop to zero, as the variant is unable to defeat monsters without attack capabilities to reduce monster's health.
Fourth, moving module ensures consistent combat presence.
 Its absence significantly reduces Success Rate by leaving the characters out of combat zones without the ability to re-enter the zone, leading to task failure.
Fifth, dodging module acts as the core survivability mechanism.
 The exclusion of it negatively impacts Success Rate because characters are defeated before they can execute full combat combo.
Sixth, handling module provides the necessary logic for error recovery.
 Its removal reduces Success Rate by leaving the characters in stuck states.
 Without the capacity to recover swiftly from these states, the variant remains unresponsive or keeps out of the combat zone, leading to a defeat or an exit from the combat.
Damage Dealt results for the ablation study are provided in \ref{app:rq4}.

\section{Discussion}

\subsection{Answering the research questions}

\rqhead{RQ1: How effective and efficient is \tool in combat?}
\tool is effective: it achieves a Success Rate of 1.0 on easy, medium, and hard tasks in Genshin Impact, with Damage Dealt reaching the maximum on these tasks, surpassing the strongest baseline by up to 0.80 on hard tasks.
\tool is also efficient: its Combat Duty Cycle remains 1.0 across all tasks, exceeding UI-TARS by 0.41 to 0.60, which indicates that the measured combat loop continues through model inference.
Although \tool does not complete very hard tasks, it outperforms all baselines on Damage Dealt by 0.14 to 0.26, suggesting that the current bottleneck is survival under boss attack patterns rather than damage output.

\paragraph{Answer to RQ1} \tool effectively executes combat tasks with various difficulty in Genshin Impact, demonstrating strong performance in terms of effectiveness and efficiency.

\rqhead{RQ2: How human-like is \tool's combat behavior, and how does it affect the mobile device?}
\tool's combat behavior is the most human-like among all evaluated agents: it attains an average human-likeness score of 2.14, close to human players (2.39) and far above the baselines (1.11 to 1.25), and wins the highest pairwise preference among agents.
Its impact on the device is close to manual play, with a negligible FPS reduction of 0.03 and slight temperature increases of 0.7 to 5.27.
These results confirm that \tool generates realistic combat workloads suitable for automated mobile game quality assurance.

\paragraph{Answer to RQ2} \tool behaves the most similarly to human players among all evaluated agents and has negligible performance impact on mobile device, validating its capacity for simulating real-world mobile game quality assurance.

\rqhead{RQ3: How well does \tool generalize?}
\tool generalizes to both Wuthering Waves and Arknights: Endfield without game-specific re-engineering of its pipeline, consistently outperforming the four baselines in Success Rate and Damage Dealt.
On easy and medium tasks in Wuthering Waves, \tool achieves a perfect Success Rate of 1.0 and the maximum Damage Dealt of 1.00, and it remains ahead of all baselines on the hard task with a Success Rate of 0.4 and Damage Dealt of 0.80.
On Arknights: Endfield, \tool attains a Success Rate of 1.0 and the maximum Damage Dealt of 1.00 across easy, medium, and hard tasks.
On the very hard tasks where no method succeeds, \tool still attains the highest Damage Dealt (0.51 and 0.60), indicating that its advantage is preserved under distribution shift.
This transferability stems from the two-phase, modular design: the LLM planner adapts the combat strategy to new monsters and character pools through external knowledge, while the five execution modules mirror the combat structure shared across RPGs.

\paragraph{Answer to RQ3} \tool generalizes across games: it maintains high effectiveness on easy, medium, and hard tasks in both Wuthering Waves and Arknights: Endfield and outperforms all baselines in Success Rate and Damage Dealt.

\rqhead{RQ4: How do the modules of \tool contribute?}
Each module contributes a distinct and necessary capability.
Removing the attacking module reduces the Success Rate to zero, confirming that attack execution is the primary driver of combat completion.
The tracking module acts as the shared perception backbone, and its removal prevents the harness from choosing appropriate actions.
The moving, dodging, and handling modules respectively ensure combat-zone presence, survivability under monster attacks, and recovery from abnormal states.
The ablation confirms that combat effectiveness emerges from the coordinated composition of specialized modules rather than any single component.

\paragraph{Answer to RQ4} Each of the five modules in \tool contributes to effectively executing combat, improving the ability to deal damage and to complete combat tasks.

\subsection{Positioning with respect to prior work}

Intrusive game testing systems such as Wuji~\citep{wuji} rely on internal game APIs and typically run on servers rather than on real devices, so they cannot generate realistic combat workloads or reflect true on-device performance; \tool instead performs quality assurance non-intrusively on the actual device through the same visual stream and touch events as human players.
Compared with script-based automation such as BetterGI~\citep{bgi}, which executes fixed, human-engineered action sequences, \tool handles dynamic combat situations that fixed scripts do not cover, such as complex multi-directional attacks and abnormal states.
Compared with vision-based reinforcement learning agents such as AI-Wukong~\citep{bmwai}, which must learn the overall combat strategy over the full action space, too complex a task for trial-and-error learning to acquire an effective policy, \tool obtains its combat strategy from pre-combat LLM planning with external knowledge and confines learning to a small dodging model for a single sub-skill.
Compared with GUI agents such as UI-TARS~\citep{uitars}, whose large-model inference latency requires pausing the game during inference, \tool confines expensive reasoning to the pre-combat phase and therefore keeps the measured combat loop free of inference latency.
Compared with vision-action foundation models such as NitroGen~\citep{nitrogen}, which are trained uniformly across many games as generalists and therefore lack game-specific understanding of skill semantics and rotations, \tool acquires game-specific strategies through the pre-combat LLM planner and executes the planned combo precisely with lightweight modules.

\subsection{Implications for designing non-intrusive game agents}

Beyond the specific system, our results suggest four design principles for agents that must operate through the same interface as human players, particularly in measurement-sensitive settings such as game quality assurance.

\paragraph{Separate knowledge-intensive planning from the latency-critical loop}
Semantic reasoning about character abilities, monster attributes, and strategy knowledge is combinatorial and therefore expensive.
Performing it before the real-time loop, where latency is tolerable, keeps inference cost out of the measured workload, while a reusable high-level plan still guides in-combat behavior.
This principle explains why \tool achieves a Combat Duty Cycle of 1.0 where foundation-in-loop agents pause the game during inference.

\paragraph{Mirror player gameplay as specialized modules over a shared perception backbone}
Human players decompose combat into tracking, attacking, moving, dodging, and handling; aligning the agent architecture with this decomposition makes each module's responsibility verifiable and its failure modes diagnosable.
A single shared perception backbone (the tracking module) feeds all downstream modules, avoiding duplicated perception cost and inconsistent world views.
This principle also underlies the cross-game generalization observed in RQ3: the modules automate the combat structure shared across RPGs, while game-specific semantics are confined to the pre-combat planner.

\paragraph{Reserve learned models for what rules cannot express}
Handcrafted rules cover common, structured behaviors cheaply and reliably; learned models should be reserved for behaviors whose state space defeats rules, such as complex multi-directional evasion.
\tool therefore uses rule-based lateral evasion for simple attacks and a small VLA model only for complex dodging, which keeps the common path lightweight without sacrificing capability on hard cases.

\paragraph{Treat non-intrusiveness as a first-class constraint}
If a harness modifies the device or the game, its measurements no longer reflect real gameplay.
Operating through the standard visual stream and touch events, with no API, hook, or device modification, makes the harness itself invisible to the measurement, which is what allows \tool's device performance metrics to remain close to manual play.

\subsection{Limitations and future works}
\label{sec:limitations}

\tool demonstrates strong performance across easy, medium, and hard combat tasks, yet consistently fails to complete the very hard tasks, such as Cocijo and Sappho Amidst the Waves in Genshin Impact, Dragon of Dirge Diff.~IV in Wuthering Waves, and Triaggelos in Arknights: Endfield.
These failures expose the limitation of insufficient survival capability for very hard boss attacks.
Future work should explore VLA architectures with greater visual reasoning capacity and investigate reinforcement learning that allows the dodging policy to adapt to attack patterns during combat and maintain survival.

\section{Conclusion}

In this paper, we present \tool, a non-intrusive, two-phase combat harness for automated mobile role-play game quality assurance that achieves precise control with low latency.
The core idea is to place expensive semantic reasoning before combat and reserve the real-time loop for lightweight modules capable of precise control, while operating games through the visual game stream and standard touch events.
By using LLM planning for character selection and combo design, and using modules for tracking, attacking, moving, dodging, and handling, \tool addresses the key requirements: adaptability on different monsters and games, real-time response, precise combat control, and low mobile-device interference with non-intrusive.
Experiments on Genshin Impact, Wuthering Waves, and Arknights: Endfield show that \tool can generate effective combat workloads while satisfying practical automated mobile game quality assurance requirements.
\tool improves combat success, damage output, and combat duty cycle over representative baselines, behaves more similarly to human players than all evaluated agents, introduces only marginal device overhead compared with manual play, and transfers to two additional RPG environments.

\appendix
\section{Combat Tasks Defined in Genshin Impact, Wuthering Waves, and Arknights: Endfield}
\label{app:task}
\setcounter{table}{0}

We define 10 combat tasks in Genshin Impact, 4 combat tasks in Wuthering Waves, and 4 combat tasks in Arknights: Endfield based on their difficulty levels, categorizing them into four distinct levels: easy, medium, hard, and very hard, as shown in Table~\ref{tab:task}.

\begin{table}[H]
\centering
\caption{Combat tasks defined in Genshin Impact, Wuthering Waves, and Arknights: Endfield by difficulty.}
\label{tab:task}
\footnotesize
\setlength{\tabcolsep}{4 pt}
\begin{tabular}{@{}lll@{}}

\toprule
\textbf{Game} & \textbf{Difficulty} & \textbf{Target Monsters} \\ \midrule
\multirow{10}{*}{Genshin Impact} & \multirow{2}{*}{Easy} & Hilichurl Fighter (HF) \\
 &  & Cryo Whopperflower (CW) \\ \cmidrule(l){2-3}
 & \multirow{3}{*}{Medium} & Blazing Axe Mitachurl (BAM) \\
 &  & Fatui Pyro Agent (FPA) \\
 &  & Ruin Drake: Earthguard (RDE) \\ \cmidrule(l){2-3}
 & \multirow{3}{*}{Hard} & Electro Hypostasis (EH) \\
 &  & Gluttonous Yumkasaur Mountain King (GYMK) \\
 &  & Goldflame Qucusaur Tyrant (GQT) \\ \cmidrule(l){2-3}
 & \multirow{2}{*}{Very Hard} & Cocijo (CO) \\
 &  & Sappho Amidst the Waves (SAW) \\ \midrule
\multirow{4}{*}{Wuthering Waves} & Easy & Fallacy of No Return Diff. I (FNR) \\
 & Medium & Sentry Construct Diff. II (SC) \\
 & Hard & Hecate Diff. III (HE) \\
 & Very Hard & Dragon of Dirge Diff. IV (DOD) \\ \midrule
\multirow{4}{*}{Arknights: Endfield} & Easy & Bonekrusher Vanguard (BV) \\
 & Medium & Quillbeast (QB) \\
 & Hard & Blitzcrash Blightshade (BB) \\
 & Very Hard & Triaggelos (TG) \\ \bottomrule

\end{tabular}
\end{table}

\section{Reward Function in PPO Training}
\label{app:reward}
\setcounter{table}{0}

We summarize the detailed reward function used in PPO training of VLA model in dodging module as Table~\ref{tab:rewards}.

\begin{table}[H]
\centering
\caption{Definition of Reward Function Components}
\label{tab:rewards}
\footnotesize
\setlength{\tabcolsep}{2.5 pt}
\begin{tabular}{@{}llc@{}}

\toprule
\textbf{Category} & \textbf{Condition} & \textbf{Reward Value ($r$)} \\
\midrule
\multirow{2}{*}{Health Status After Dodging} & Health maintained & $+0.1$ \\
 & Health decreased & $-3.0 \times \Delta HP_{\%}$ \\
\midrule
\multirow{2}{*}{Position Change After Dodging} & Approaching center & $\min(0.2, 0.02 \times d_{near})$ \\
 & Moving away from center & $\max(-2.0, -0.05 \times d_{far})$ \\
\midrule
\multirow{2}{*}{Position Change During Dodging} & Distance away $\leq 5$ & $+0.05$ \\
 & Distance away $> 5$ & $\max(-1.0, -0.025 \times d_{far})$ \\
\midrule
\multirow{3}{*}{Model Output Format} & Correct format & $+0.1$ \\
 & Incorrect format & $-0.5$ \\
 & Unparseable error & $-6.5$ \\
\bottomrule

\end{tabular}
\end{table}

\section{Dataset Construction}
\label{app:dataset}

For YOLO models, we extract 1000 to 2750 frames per monster from gameplay recordings and annotate each frame using makesense.ai~\citep{makesenseai} to label monster bounding boxes and state categories.
The dataset is split 8:1:1 for train, valid, and test.
For orientation recognition, we annotate 515 minimap samples for character orientation and camera direction. Random edge cropping augments these to 2575 images.
For VLA in SFT stage, we record 63 gameplay videos of continuous dodge sequences and manually annotate the dodge direction for each clip. Sets of four consecutive frames paired with their direction label form one data pair; 1057 such pairs are generated in total for VLA fine-tuning.

\section{Details of Baselines and Full Justifications for Excluded}
\label{app:baselines}

We compare our method with approaches including BetterGI~\citep{bgi}, UI-TARS~\citep{uitars}, AI-Wukong~\citep{bmwai}, and NitroGen~\citep{nitrogen}:

\begin{itemize}
    \item BetterGI~\citep{bgi} represents script-based game combat automation. It executes a fixed, human-engineered loop consisting of predefined action sequences in game to combat, relying on hard-coded timing and deterministic control logic.
    \item AI-Wukong~\citep{bmwai} represents vision-based reinforcement learning approach in game combat. It utilizes a pre-trained ResNet as the visual backbone to extract high-level spatial features, then feed features into a Deep Q-Network for training a combat agent.
    \item UI-TARS~\citep{uitars} represents screenshot-based GUI agents trained from large-scale interaction data. It develops a GUI agent by fine-tuning large vision-language models on extensive interaction datasets, employing an end-to-end approach to directly map screen pixels to actions.
    \item NitroGen~\citep{nitrogen} represents transformer-based vision-action foundation models. It employs a Vision Transformer encoder to process multi-modal visual inputs, coupled with a Diffusion Transformer backbone for high-fidelity action latent prediction.
\end{itemize}

To facilitate a fair comparison, we adapted the selected baselines to the mobile RPG environment:
(1) For BetterGI~\citep{bgi}, which is originally designed for Genshin Impact on PC environments, we re-implement the automated combat module in Python to align with our framework and using our framework to operate Genshin Impact, Wuthering Waves, and Arknights: Endfield on the mobile device;
(2) For AI-Wukong~\citep{bmwai}, which is originally designed for Black Myth: Wukong on PC environments, we integrate our framework into it for mobile screen captures and action execution, and remove game-specific mechanisms originally designed for Black Myth: Wukong to maintain generalizability. Specifically, health point recognition the reward mechanism relies on is recalibrated to account for the distinct UI layouts and visual characteristics of Genshin Impact, Wuthering Waves, and Arknights: Endfield. Given the dynamic and irregular positioning of monster health point bars in Genshin Impact, which poses challenges for automated detection, we employed a human-in-the-loop approach: health point fluctuations were manually monitored and fed into the system to ensure the accuracy of reward signals during training;
(3) UI-TARS~\citep{uitars} is deployed via the UITARS\_MobileAgent~\citep{uitarsmobile} library, with prompt engineering specifically optimized to handle the Genshin Impact, Wuthering Waves, and Arknights: Endfield combat.  To address the significant inference latency inherent in the model which affects the game capability, we implement a synchronized execution protocol: the game environment is paused upon the initiation of an inference request and resumed after the model returns the corresponding action commands;
(4) For NitroGen~\citep{nitrogen}, we develop a control mapping layer that translates standard gamepad inputs into mobile touch events, based on the default controller schemas of Genshin Impact, Wuthering Waves, and Arknights: Endfield, and use our framework for screen captures and action execution.

Several methods are excluded from our experiments for the following reasons: (1) We exclude TiG~\citep{tig} as it is designed for global strategy generation and not applicable to real game combat. (2) We exclude Wuji~\citep{wuji} as it runs on servers instead of game devices. (3) We exclude VistaWise~\citep{vistawise}, Jarvis-VLA~\citep{jarvis-vla}, Optimus-2~\citep{optimus2}, and OpenHA~\citep{openha} since they are designed for other types of games and not applicable to RPG game combat. (4) We exclude Cradle~\citep{cradle} as it is designed for general-purpose computer control and lacks support for mobile environments. (5) We exclude VARP~\citep{varp}, CombatVLA~\citep{combatvla}, Game-TARS~\citep{gametars}, TITAN~\citep{titan}, PORTAL~\citep{portal} SIMA 2~\citep{sima2}, and Lumine~\citep{lumine}, because they are not open-source.

\section{Hyperparameter Settings}
\label{app:hyper}

YOLOv11-X model for each monster is fine-tuned over 200 epochs with batch size 16, input resolution 640 $\times$ 640, and weights of 7.5 for box loss, 1.5 for classification loss, and 3.0 for distribution focal losses.
All remaining settings follow Ultralytics defaults~\citep{yolo}.
The orientation recognition CNN is trained with the default parameters of the pytorch\_image\_regression framework~\citep{pytorch_image_regression}.
Qwen2.5-VL-3B is fine-tuned in the SFT stage at a learning rate of 2e-7, batch size 1, for 1 epoch.
PPO training is subsequently run on monsters Cocijo and Sappho Amidst the Waves until convergence with 500 to 1000 steps, using the algorithm adapted from trl library~\citep{trl}.

\section{Details of Experiments Metrics}
\label{app:metrics}

We discuss experiments metrics used in experiments as below:

\begin{itemize}
    \item \textit{Success Rate} is defined as the ratio of successfully completed combat missions, where a trial is marked as a success if the target enemy is defeated, to the total number of evaluation trials. A higher Success Rate indicates that the harness can consistently achieve defeating all enemies across varied combat scenarios. It provides a high-level view of the effectiveness of harness in combat.
    \item \textit{Damage Dealt} is total damage output produced by the harness throughout the entire duration of a single combat encounter. A higher Damage Dealt proves the harness can eventually overcome a monster's health. It serves as an indicator of total combat capacity.
    \item \textit{Combat Duty Cycle} is defined as the ratio of the time the harness spends executing combat to the total session duration, which includes the overhead of model inference and decision-making processes. A high Combat Duty Cycle indicates an efficient decision-making pipeline, where the harness spends the majority of its time interacting within the combat rather than waiting for model computations. It captures the temporal efficiency of the harness's internal architecture.
    \item \textit{Frames Per Second} represents the temporal fluidity of the game rendering. When the FPS of our method is comparable to that of manual playing, it indicates that the harness does not introduce significant computational latency or bottleneck the rendering pipeline, ensuring the game's native performance.
    \item \textit{CPU Temperature} reflects the computational load. A stable CPU temperature relative to manual play suggests that \tool minimizes additional on-device computation, ensuring that the automated combat process has negligible interference with the device's overall performance profile.
    \item \textit{GPU Temperature} indicates the intensity of graphical processing. Maintaining temperatures within a range similar to human play demonstrates that \tool requires minimal mobile graphic resources, thereby allowing for a more accurate simulation of real-world gaming performance.
    \item \textit{Battery Temperature} indicates power consumption. Maintaining a battery temperature similar to human play indicates that the method has a minimal impact on power draw, which is essential for replicating the thermal and energy behavior of a real user environment.
\end{itemize}

\section{Damage Dealt Results of Ablation Study}
\label{app:rq4}
\setcounter{table}{0}

Table~\ref{tab:rq4_dd_cd} presents the Damage Dealt results of \tool and its variants for the ablation study tasks in Genshin Impact.
These results complement the Success Rate analysis in Section~\ref{sec:rq4} and further quantify each module's contribution to total damage output.
First, removing the tracking module causes a marked decline in Damage Dealt across hard and very hard tasks, confirming that the inability to perceive the monster's state reduces both attack precision and survival duration.
Second, removing the attacking module causes Damage Dealt to drop to zero across all tasks, as no attacks are issued to reduce the monster's health.
Third, the absence of the moving module reduces Damage Dealt by displacing characters out of effective combat range, preventing attacks from landing.
Fourth, excluding the dodging module lowers Damage Dealt in hard and very hard tasks, as characters are defeated before completing their planned combat combos, thereby cutting total damage short.
Fifth, removing the handling module reduces Damage Dealt by leaving characters in abnormal states such as climbing, freezing, or swimming, which disrupts sustained damage output.
These observations are consistent with the Success Rate findings and collectively confirm that each execution module contributes an indispensable role in maximizing total combat damage.

\begin{table}[H]
\centering
\caption{Statistical results of Damage Dealt of \tool and its variants for combat tasks in Genshin Impact. Results are reported in the format of Mean (Standard Deviation).
For the Damage Dealt metric, values are normalized such that a full monster health bar equals 1.0.
Bold text highlights the most superior results in each row.}
\label{tab:rq4_dd_cd}
\scriptsize
\setlength{\tabcolsep}{1.5 pt}
\begin{tabular}{@{}ccccccc@{}}

\toprule
\textbf{Task} & \textbf{\tool} & \textbf{w/o Tracking} & \textbf{w/o Attacking} & \textbf{w/o Moving} & \textbf{w/o Dodging} & \textbf{w/o Handling} \\
\cmidrule(lr){1-7}
HF & \textbf{1.00(0.00)} & \textbf{1.00(0.00)} & 0.00(0.00) & \textbf{1.00(0.00)} & \textbf{1.00(0.00)} & \textbf{1.00(0.00)} \\
BAM & \textbf{1.00(0.00)} & \textbf{1.00(0.00)} & 0.00(0.00) & \textbf{1.00(0.00)} & \textbf{1.00(0.00)} & \textbf{1.00(0.00)} \\
EH & \textbf{1.00(0.00)} & 0.82(0.15) & 0.00(0.00) & 0.85(0.16) & 0.90(0.16) & 0.71(0.32) \\
CO & \textbf{0.26(0.13)} & 0.06(0.04) & 0.00(0.00) & 0.08(0.03) & 0.06(0.04) & 0.07(0.05) \\
\bottomrule

\end{tabular}
\end{table}

\section*{Declaration of generative AI and AI-assisted technologies in the manuscript preparation process}
During the preparation of this work the authors used Kimi in order to improve the language and readability of the manuscript.
After using this tool, the authors reviewed and edited the content as needed and take full responsibility for the content of this publication.

\section*{Acknowledgements}
This work was supported by the National Natural Science Foundation of China (Grant No.~62232014 and Grant No.~62402214), the China Postdoctoral Science Foundation (Grant No.~2025T180420 and Grant No.~2025M781473), and the Jiangsu Funding Program (Grant No.~2025ZB317).

\bibliographystyle{elsarticle-harv}
\bibliography{references}

\end{document}